\documentclass[aps,prl,twocolumn,groupedaddress]{revtex4}
\usepackage{graphicx,psfrag}
\newcommand{\be}{\begin{equation}}

\newcommand{\ee}{\end{equation}}
\newcommand{\bea}{\begin{eqnarray}}
\newcommand{\eea}{\end{eqnarray}}
\newcommand{\ba}[1]{\begin{array}{*{#1}{c}}}
\newcommand{\ea}{\end{array}}

\begin{document}

\title{Relaxation times of kinetically constrained spin models with glassy dynamics}

\author{Nicoletta Cancrini$^{1}$}

\author{Fabio Martinelli$^2$}

\author{Cyril Roberto$^3$}

\author{Cristina Toninelli$^4$}
\email[]{cristina@corto.lpt.ens.fr}
\affiliation{$^1$Dip.Matematica Univ. l'Aquila, 1-67100 L'Aquila, ITALY
  and INFM unit\'a Roma La Sapienza, \\
$^2$Dip.Matematica, Univ. Roma Tre, Largo S.L.Murialdo 00146,
  Roma, ITALY, \\
$^3$L.A.M.A., Univ Marne-la-Vall\'ee, 5 bd Descartes
  67454 Marne-la-Vall\'ee FRANCE,\\
$^4$ I.H.E.S., Les Bois Marie 35, route de Chartres F-91440,
Bures-sur-Yvette, FRANCE 
}
\begin{abstract}

We analyze the density and size dependence of the relaxation time
$\tau$ for kinetically constrained spin systems.  These have been
proposed as models for strong or fragile glasses and for systems
undergoing jamming transitions. For the one (FA1f) or two (FA2f) spin
facilitated Fredrickson-Andersen model at any density $\rho<1$ and for
the Knight model below the critical density at which the glass
transition occurs, we show that  the
persistence and the spin-spin time auto-correlation functions decay
exponentially. This excludes the stretched exponential relaxation
which was derived by numerical simulations. 
 For FA2f in $d\geq 2$, we also prove a super-Arrhenius
scaling of the form $\exp(1/(1-\rho))\leq \tau\leq\exp(1/(1-\rho)^2)$.
For FA1f in $d$=$1,2$ we rigorously prove the power law scalings
recently derived in \cite{JMS} while in $d\geq 3$ we obtain upper and
lower bounds consistent with findings therein.  Our results are based
on a novel multi-scale approach which allows to analyze $\tau$ in
presence of kinetic constraints and to connect time-scales and
dynamical heterogeneities. The techniques
are flexible enough to allow a variety of constraints and can also be
applied to conservative stochastic lattice gases in presence of
kinetic constraints.
\end{abstract}
 \keywords{} \maketitle Despite a great deal of theoretical and
experimental efforts, the main issues in understanding {\sl
liquid--glass} or {\sl jamming transitions} remain
unsolved.  These phenomena occur in different systems: supercooled
liquids, colloidal suspensions, vibrated granular
materials~\cite{liquids,colloids}.  Characteristic features of
a {\sl glassy behavior} include dramatic slowing down of dynamics when
a proper external parameter is tuned (e.g.  temperature for liquids)
and the occurrence of a complex, non exponential and spatially
heterogeneous \cite{hetexp} relaxation process. Experiments show that,
if an {\sl ideal glass transition} occurs, it exhibits mixed first and
second order features. Indeed, even if the divergence of the
relaxation time signals a second order like transition, no static
diverging correlation length is detected. Furthermore relaxation times
for fragile liquids diverge in a super-Arrhenius way: $\tau\simeq
\exp\left(A(T-T_c)/T\right)$ with $A(x)\uparrow \infty$ as
$x\downarrow 0$, a signal of a cooperative relaxation on increasingly
large scales as $T$ decreases.  \\ \indent A class of microscopic
models that have been proposed in the attempt of understanding these
phenomena are the so-called {\sl kinetically constrained models} (KCM)
(\cite{RS,FA,JE,GB,SE,AD,GC,WBG,BG,JMS,R,FB,GPG,H,F,BH,RJM,noiKA,letterTBF}
and references therein). KCM are systems of particles on a discrete
lattice with no static interactions beyond hard core and evolution is
given by a Markovian stochastic dynamics. The elementary move is
either a jump of a particle to a nearby empty site in the conservative
(Kawasaki) case or the creation/destruction of a particle in the non
conservative (Glauber) case.  In both situations the associated rates,
if positive, verify detailed balance with respect to a simple product
measure. However, and this is the most interesting feature, a move at
site $x$ can occur only if the configuration around $x$ satisfies
certain constraints.  The latter mimic the local constraints that may
occur in the physical systems and which may cause the dynamical
arrest. Indeed, for a proper choice of the constraints, KCM show
glassy features including stretched exponential relaxation,
super-Arrhenius slowing down and dynamical heterogeneities. Several
works and a great deal of numerical simulations have recently been
devoted to understand the mechanism inducing these glassy properties
and to derive the typical time-scales as well as the asymptotic form
of correlation functions
(\cite{RS,FA,JE,GB,SE,AD,GC,WBG,BG,JMS,R,FB,GPG,H,F,BH,RJM,noiKA,letterTBF}).
Numerical simulations are however very delicate because of the rapid
divergence of $\tau$ as the density is increased. Furthermore, finite
size effects often have non trivial scalings.  In this work we present
a new general probabilistic technique through which we obtain rigorous
results on the dependence of $\tau$ on the size of the lattice and on
the particle density, $\rho$. As a by-product we also obtain
meaningful bounds on the long-time behavior of time correlations and
persistence function.\\ \indent We first introduce the models and our
results and then sketch the technique, referring to \cite{CMRT} for
rigorous proofs.  We focus on non conservative models, also known as
{\sl kinetically constrained spin models} (KCSM) or {\sl facilitated
spin models}.  Conservative models will be analyzed elsewhere
\cite{CMRT}.  KCSM are defined as follows. Each site $x$ can be
occupied, $\eta_x$=$1$, or empty, $\eta_x$=$0$, and it changes its
current state with rate
$\left[(1-\rho)\eta_x+\rho(1-\eta_x)\right]f_x(\eta)$.  Here
$f_x(\eta)$ does not depend on $\eta_x$ and it encodes the kinetic
constraint. Thus detailed balance is satisfied w.r.t. the Bernoulli
product measure, $\mu_{\rho}$, at density $\rho$.  KCSM can be divided
into two classes: (i) non-cooperative and (ii) cooperative.  In a
non-cooperative model it is possible to completely empty {\sl any}
configuration containing somewhere a finite seed of vacancies. In a
non-cooperative one that is not the case.  The class (ii) can be
further divided into: (iia) models that are ergodic in the
thermodynamic limit at any $\rho<1$; (iib) models that display an
ergodicity breaking transition at a critical density $\rho_c<1$. For
the latter, above $\rho_c$, there exists an infinite spanning cluster
of particles mutually blocked by the constraints. Finally (iib) can be
classified as discontinuous or continuous according to the character
of the percolation transition of the blocked structure.

Among non cooperative models we consider the one spin facilitated
Fredrickson-Andersen model \cite{FA} (FA1f), which recently received a
renewed attention as a model for strong glasses \cite{WBG}.  A move is
allowed only if at least one of the nearest neighbors is empty:
$f_x(\eta)$=$1$ if $~\sum_{y~ n.n. x} (1-\eta_x)>0$, $f_x(\eta)$=$0$
otherwise.  In \cite{WBG} a dynamical field theory was derived which
gives ($q$=$1$$-$$\rho$): $\tau\propto 1/q^z$ with $z$$=$$3$ for
$d$$=$$1$, $z$$=$$2+\epsilon(d)$ with $\epsilon(2)\simeq 0.3$,
$\epsilon(3)\simeq 0.1$ and $\epsilon(d\geq 4)=0$. A recent exact
mapping into a diffusion limited aggregation model, its
renormalization and a careful treatment of the symmetries involved
\cite{JMS} gives instead $d$$=$$2$ as the upper critical dimension and
$\epsilon(d)$=$0$ in $d\geq 2$.

Our results for FA1f 
are the followings. If $\tau$ denotes the inverse of the
spectral gap of the Liouvillian operator generating the stochastic
dynamics, i.e. the worst relaxation time on all one time quantities, we
get: $\tau\propto 1/q^3$ in $d=1$, $1/q^2<\tau\leq 1/(q^2\log 1/q)$ in
$d=2$ and $1/q^{1+2/d}<\tau\leq 1/q^2$ in $d=3$. 
These rigorous results lead to
$\epsilon(2)=0$ and $\epsilon(3)\leq 0$, disproving the findings in 
\cite{WBG} and confirming those in \cite{JMS}. 
Moreover our method allows us to
identify explicitly the slowest modes which dominate relaxation.
Furthermore, it is easily adapted to
{\sl any} possible choice of non cooperative constraints, e.g. to
models in which more than one spin is needed to empty the whole
lattice.
We can also treat models with asymmetric (e.g. East model
\cite{JE}) or partially 
asymmetric constraints (\cite{GB}) which have been proposed
to model fragile glasses \cite{GC} and the strong/fragile crossover. 
In particular for East, we obtain 
$\log\tau\simeq(\log(1/q))^2$ as in \cite{SE,AD}.  
Finally 
we can deal with the
persistence function
$$
F(t)=\int\,\,
d\mu_\rho(\eta({0}))\ P[\eta_0(s)=\eta_0({0})\ \forall s<t] \ .
$$
In great generality \cite{CMRT} we prove that whenever the global
relaxation time $\tau$ is finite, $F(t)$ decays exponentially fast. In
particular, for FA1f we obtain $\tau_F\le q^{-1}\tau$ for small $q$,
where $\tau_F$ is defined via $F(\tau_F)=e^{-1}$. Although these
findings disprove any stretched exponential decay at large times, for
FA1f in $d=1$ our bound $\tau_F \leq 1/q^4$ does not preclude the
stretched form \cite{BG,WBG} $F(t)\simeq \exp(-\sqrt{q^3t})$ when
$q\downarrow 0$ and $q^3 t\simeq O(1)$. This is indeed the regime
where numerical simulations are performed \cite{BG}.

Among cooperative models without transition (iia) we consider the two
spin facilitated FA model (FA2f) on square and cubic lattices.  Here the
constraint requires that at least two of the surrounding sites are empty
in order for the rate to be non zero.  Originally \cite{FA} a dynamical transition
at a finite density was predicted but later it was shown that
it cannot occur \cite{R,FB,AL}.  Still, FA2f 
is relevant for fragile glasses since both super-Arrhenius divergence
for $\tau$ and stretched exponential relaxation
have been detected \cite{FA,R,FB,GPG,H,BH,F}.  Indeed, both the
spin-spin time autocorrelation and the persistence function
are fitted with $\exp\left(-(t/\tau)^{\beta}\right)$ with $\beta$ decreasing
as the density $\rho$ is increased \cite{GPG,F,H}. 
As pointed out in \cite{RS}, beyond the general recognition that
FA2f behaves like fragile glasses, little is known about the scaling of
$\tau$. Different fits have been proposed: Adam-Gibbs form \cite{FB},
Vogel-Fulcher form \cite{GPG} and $\exp(c/q)$ \cite{BH}.  The latter,
which corresponds to a super-Arrhenius form when $q$ is rewritten in
terms of the temperature, is supported by the conjecture that relaxation
occurs via the diffusion of critical droplets of size $1/q$
over distances $\exp(c/q)$ \cite{R}.

Our results for FA2f in $d=2$ and $d=3$ are the followings.  We
prove that $~\exp(1/q)\leq\tau\leq \exp(1/q^2)~$
which establishes a super-Arrhenius scaling and excludes
the Vogel Fulcher form \cite{GPG}.
Furthermore, since our results hold uniformly on the system size, we
obtain that relaxation in infinite volume is purely exponential
at any fixed $q$. Furthermore, as for FA1f, we get
a strict bound on the crossover regime where a stretched form 
may occur if we let simultaneously $q\downarrow 0$ and $t\uparrow\infty$.
These findings
contradict the asymptotic stretched exponential relaxation in
\cite{GPG,F,H}. The
fit with a stretching exponent $\beta<1$ should be due to the rapid
divergence of $\tau$, which was also a fitting parameter. 

Among (iib) models with continuous transition, we consider the two
dimensional North East model (NE). 
For NE both the up and right neighbors
should be empty in order for a move to be allowed and  a
continuous transition occurs 
at the critical density of directed percolation,
$\rho_{dp}$ \cite{RJM}. 
Finally, for (iib) with a discontinuous transition, we consider the
recently proposed Knight model \cite{letterTBF}. 
We refer to \cite{letterTBF} for the precise definition of the constraints; we only recall that 
the transition occurs again at 
$\rho_{dp}$ and has the remarkable features 
of an {\sl ideal glass transition}. Indeed a finite
fraction of the system is frozen at $\rho_{dp}$ and the size $\xi$ of
blocked structures diverges faster than any power law as $\rho\uparrow
\rho_{dp}$. This leads to a super-Arrhenius divergence  
if $\tau\simeq \xi^z$ is assumed.

Our main results for NE and Knights are the followings. 
For $\rho<\rho_{dp}$ we identify a constant which bounds $\tau$ 
uniformly on the system size, $L$.
Therefore,
even if an ergodicity breaking transition occurs,
relaxation is purely exponential in the ergodic region.
Moreover, we get that $\tau$ diverges when $\rho\nearrow\rho_{dp}$ and
$\tau\propto\exp(L~c(\rho))$ for $\rho>\rho_{dp}$.
This result provides a
possible test on the validity of Knights for systems undergoing
jamming transitions.

\begin{figure}[t]
\psfrag{Lde}[][]{$L^{\delta}$}
\psfrag{b}[][]{b)}
\psfrag{a}[][]{a)}
\psfrag{L}[][]{$L$}
\psfrag{Lc}[][]{$\delta L_c$}
\psfrag{1}[][]{$1$}
\psfrag{2}[][]{$2$}
\psfrag{3}[][]{$3$}
\psfrag{4}[][]{$4$}
\centerline{
  \includegraphics[width=0.45\columnwidth]{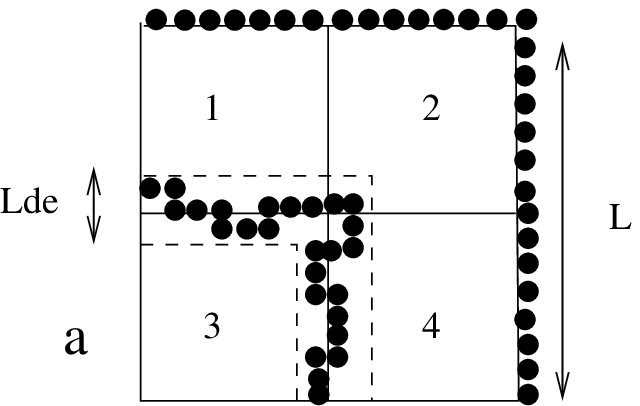}
\hspace{0.3cm}
  \includegraphics[width=0.44\columnwidth]{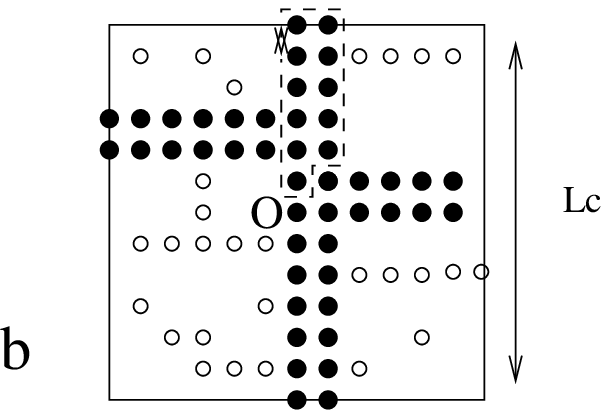}
}
\caption{a) Block dynamics for GM: percolating path of good sites
  ($\bullet$) required to renew configuration on $3$. b) Blocking event
  for FA2f. $\bullet$ ($\circ$) stand for particles which do (do not)
  belong to the backbone.  Sites inside dotted line form one of the
  sequences of $\geq \delta L_c/2$ 
sites to be emptied before $O$.}
\label{fig1}
\end{figure}

Let us now sketch the main ideas beyond our approach, rigorous proofs
will be reported elsewhere \cite{CMRT}. Here we focus on $d=2$.
Results in $d=1$ can be derived analogously; we will comment at the
end on $d=3$.  We first introduce a new model referred to as the {\sl
General Model (GM)} and derive results for its relaxation time,
$\tau_{\tiny{GM}}$. Then we show how to map a given KCSM into GM by a
renormalization technique.  GM can be described as follows. Instead of
two-valued occupation variables consider N-valued variables, $n_x\in
S=(0,\dots N-1)$, and a probability measure $\nu$ on $S$.  We identify
a subset $G$ of $S$ which we call the {\sl good event} and we declare
{\sl good} a site $x$ if $n_x\in G$. The dynamics is defined as
follows. Each site $x$ waits a mean one exponential time and then
$n_x$ is refreshed by a new value $n'_x$ sampled from $\nu$, provided
its three North, North-East and East neighbors ({\sl i.e.}
$x+\vec e_1, x+\vec e_1+\vec e_2,x+\vec e_2$, $\vec e_i$ being the
unit basis vectors) are good. If this constraint is not satisfied,
$n_x$ remains unchanged.  We consider GM on a square lattice
$\Lambda_L$ of linear size $L$ with good boundary conditions on the
top and right boundaries to ensure ergodicity. In order to evaluate
$\tau_{GM}(L)$ we follow the {\sl bisection method}
\cite{Martinelli}. Partition $\Lambda_{L}$ into four blocks as in
fig.\ref{fig1}a) and define the following auxiliary accelerated
dynamics. Each block waits a mean one exponential time and then its
configuration is replaced by a new one chosen according to the product
equilibrium probability given by $\nu$. On the top right block (block
$2$ in fig.\ref{fig1}a)) this move is always allowed because of the
boundary conditions. For the others, a constraint should be satisfied:
on an l-shaped frame of width $L^\delta$, $\delta<1$, there should be
a percolating cluster of good sites in the current configuration as in
fig.\ref{fig1}a).
In other words, the constraint requires the good GM boundary
conditions on the blocks $1,3,4$ (on $2$ they are automatically
guaranteed). The relaxation time $\tau_{GM}(L)$ is then bounded by
the product
$$
\tau_{GM}(L)\leq \tau_{\rm block}(L)\,\tau_{GM}(L/2)
\nonumber
$$
 with $\tau_{\rm block}(L)$ the relaxation time for the block
dynamics \footnote{Here we are neglecting some unessential extra pre-factors
  arising from the overlapping of the $L^\delta$ boundary of a block
  with the other blocks, see \cite{CMRT} for a proper analysis.}.
  The above inequality
corresponds intuitively to a two steps relaxation: first on the block
scale, then inside each individual block. Notice that, trivially,
$\tau_{GM}(L)$ is smaller than the worst case over the boundary
conditions of the relaxation time on scale $L/2$.  However, such
inequality is useless because without the good b.c. GM dynamics is not
ergodic (i.e.  $\tau$ is infinite).  In other words, working with a
constrained block dynamics is essential. Then, by dividing
$\Lambda_{L/2}$ into four blocks and so on up to constant size, we get
$$
\tau_{GM}(L)\leq c\prod_n\tau_{\rm block}(2^{-n}L),
\nonumber
$$
where $c$ is a finite constant and the product contains $O(\log L)$
terms.  Therefore we get a bound for $\tau_{GM}(L)$ which does not
dependent on $L$ provided that the product over the $\tau_{\rm
block}$'s converges. In turn this occurs if the probability that a
site is good, $p$, is larger than a finite threshold, $p_c<1$.  Indeed
one can easily show that, for $p\simeq 1$, $ \tau_{\rm block}(L)\simeq
\left(1+\exp(-mL^\delta)\right) $ where $\exp(-mL^\delta)$ comes from
the probability that the constraint for the block dynamics on scale L
is violated.

Let us now explain how to apply such a result to the KCSM models
described above. We consider for definiteness FA2f in $d=2$ with empty
boundary conditions on the top and right borders.  We say that a
region $V$ is {\sl internally spanned} for the configuration $\eta$
if, when considering all the sites outside $V$ occupied, $\eta$ can be
completely emptied using (internal) allowed moves. The probability for
$\Lambda_L$ to be internally spanned has been evaluated in the context
of bootstrap percolation: it goes to one exponentially fast when $L$
exceeds the crossover length $L_c\simeq \exp(c/q)$ with $c=\pi^2/18$
\cite{Ho}.  Divide now the square lattice into disjoint blocks of size
$k\,L_c$, $k\gg 1$, and identify each block by the coordinates of its
center.  On each of these renormalized sites we define a configuration
space $S=\{0,1\}^{N_b}$, with $N_b$ the number of sites in a block.
In $S$ we identify the good event $G$ as the set of configurations
such that the block is internally spanned.  Because of our choice of
$k$ the probability that a renormalized site is good is $\simeq 1$. We
can thus run the GM dynamics on the renormalized lattice and get
$\sup_L\,\tau_{\rm GM}(L)\simeq 1$.  Since in GM dynamics blocks are
updated only if ``enough'' surrounding are internally spanned, by the
two steps relaxation argument we get
$$
\tau(L)\leq \tau_{\rm GM}(L)\,  \tau( kL_c) \simeq \tau( kL_c).
\nonumber
$$
where $\tau(L)$ is the relaxation time for FA2f.  In the infinite
volume limit and for any $\rho<1$ this leads to an exponential
relaxation for all one time functions as well as for the persistence
function using a Feyman-Kac bound \cite{CMRT}. At the same time the
density dependence of $\tau$ is completely encoded in the relaxation
time on a lattice of size $L_c(\rho)$. To evaluate the latter we
reduce the scale from $L_c$ to $1/q^2$ via a strategy similar to the
previous one. However, on scales smaller than $L_c$ the event ``the
block is internally spanned'' becomes very unlikely and we are forced
to make a different choice for the constraints of the renormalized
dynamics in order to keep $\tau_{\rm GM}\simeq 1$. Our choice for the
good event $G$ is suggested by the following two observations: (i) any
empty segment of length $\ell$ can be displaced by one step in a given
direction if there is at least one vacancy on the adjacent segment in
that direction; (ii) the probability that there exists a fully
occupied segment of length $\ell$ inside a critical square of size
$L_c$ is very small as soon as $\ell\gg 1/q^2$. Thus, we choose good
events which force the not fully occupied condition on segments of
length $1/q^2$. By applying a bisection procedure analogous to the one
used for GM together with elementary paths arguments, we get
$$  
\tau(L_c)\leq cL_c\tau(1/q^2)
$$
Finally we bound $\tau(1/q^2)$ with the highest entropy cost to get
$\tau_{FA}(L)<cL_c\exp(1/q^2)=O(\exp(1/q^2))$.

The results for the other $2$-dimensional models (FA1f, NE and Knight)
are obtained analogously. Furthermore, for a generic choice of the
constraints, we find that $\tau$ is dominated by $\tau(L_c)$ where
$L_c$ is the crossover length for a region to be internally spanned
(with the chosen constraints). Therefore $L_c$, which can be
determined by a deterministic bootstrap-like procedure, is the
relevant size over which numerical simulations should be performed.
By analyzing the constraints one can also further reduce the problem
to a much smaller scale, as we do for FA2f.

Lower bounds for $\tau$, directly in the thermodynamic limit,
can be established either via a suitable choice of test functions in
the variational characterization of $\tau$ or using our general
result for
the persistence function: $F(t)\leq \exp(-tq/\tau)$ \cite{CMRT}.
Consider an event $B$, called the {\sl blocking event}, and let
$P_B(t)$ be the probability that the origin is occupied for all times
$s\leq t$, worst case among all the starting configuration in
$B$.  The inequality for $F(t)$ implies $\mu_{\rho}(B)P_B(t)\leq
\exp(-tq/\tau)$.  We define the blocking event $B$ as the set of
configurations for which, after standard bootstrap percolation (i.e.
all possible FA2f moves in sequence) inside $\Lambda_{\delta L_c}$,
a backbone of particles containing the origin survives. By choosing
$\delta \ll 1$ we have $\mu_\rho(B)\simeq 1$.  In infinite volume this
backbone will eventually get unblocked thanks to the vacancies outside
$\Lambda_{\delta L_c}$. However, this requires an ordered sequence of
at least $\delta L_c/2$ moves (fig.\ref{fig1}b). Thus, $P_B(t=\epsilon
\delta L_c)\simeq 1$ for sufficiently small $\epsilon$.  Therefore
$O(1)\leq \exp(-tq/\tau)$ for $t\simeq \epsilon\delta L_c$,
i.e. $\tau\geq O(L_c)$.  In $d=3$ lower bounds can also be obtained as
before: using the results in \cite{CM} for $L_c$ we get
$\tau\geq\exp(c/q)$ for FA2f and $\tau\geq\exp\exp(c/q)$ for FA3f. The
latter is in agreement with the conjecture in \cite{R} and disproves
the possibility of an Adam Gibbs form \cite{F} which corresponds to
$\tau\simeq \exp(c/q)$. Finally the sharp lower bounds mentioned
before for FA1f are obtained by a careful choice of a test
function. \\ In summary, we have developed a technique which allows to
obtain rigorous results on $\tau$ for KCSM via the knowledge of the
typical regions which have to be rearranged to perform a movement. In
other words, we have been able to connect rigorously time-scales and
dynamical heterogeneities, a subject which have recently received
great attention in the glass community \cite{hetexp}.  The main new
results established via this technique are super-Arrhenius behavior
and exponential relaxation for cooperative models. Moreover our method
suggests the typical length scale on which numerical simulations
should be carried on. As for future developments, we believe that our
technique can be applied to finite dimensional models other than KCM,
e.g. models with static interactions beyond hard core and to
conservative KCM models.\\

 We thank G.Biroli for a careful reading of the manuscript and the
IHES for its kind hospitality.

\end{document}